\documentclass[runningheads]{svmult}

\usepackage{makeidx}   % allows index generation
\usepackage{graphicx}  % standard LaTeX graphics tool
\usepackage{subeqnar}  % subnumbers individual equations
\usepackage{multicol}  % used for the two-column index
\usepackage{physprbb}  % modified textarea for proceedings,
\makeindex             % used for the subject index
\begin{document}
\title*{Predicting and generating  time series by neural networks: 
An investigation using statistical physics}

\titlerunning{Time series}
% allows abbreviation of title, if the full title is too long
% to fit in the running head
%
\author{Wolfgang Kinzel}

\institute{Institute for Theoretical Physics, University of W\"{u}rzburg,\\
 Am Hubland, 97074 W\"{u}rzburg, Germany}

\maketitle              % typesets the title of the contribution

\begin{abstract}
An overview is given about the statistical physics of neural networks
generating and analysing time series. Storage capacity, bit and
sequence generation, prediction error, antipredictable sequences,
interacting perceptrons and the application on the minority game are
discussed.  Finally, as a demonstration a perceptron predicts bit
sequences produced by human beings.
\end{abstract}

\section{Introduction}
In the last two decades there has been intensive research on the
statistical physics of neural networks 
\cite{Hertz:NeuralComp,Engel:00,Kinzel:Rev}. 
The cooperative behaviour of neurons interacting by
synaptic couplings has been investigated using mathematical models
which describe the activity of each neuron as well as the strength of
the synapses by real numbers. Simple mechanisms change the activity of
each neuron receiving signals via the synapses from many other ones,
and change the strength of each synapse according to presented
examples on which the network is trained.

In the limit of infinitely large networks and for a set of 
random examples there exist mathematical tools to calculate 
properties of the system of interacting neurons and synapses 
exactly. For many models the dynamics of the network 
receiving continuously new examples has been described by 
nonlinear ordinary differential equations for a few order 
parameters describing the state of the system \cite{BiehlC}. 
If a network is trained on the total set 
of examples, the stationary state has been described by 
a minimum of a cost function. Using methods of the statistical 
mechanics of disordered systems (spin glasses), the properties 
of the network can be described by nonlinear equations of 
a few order parameters.

It turns out that already very simple models of neural 
networks have interesting properties with respect to information 
processing. A network with $N$ neurons and $N^2$ synapses 
can store a set of order $N$ patterns simultaneously. Such a 
network functions as a content--addressable, distributed and associative memory.

Already a simple feedforward network with only one layer of synaptic
weights can learn to classify high dimensional data.  When such a
network (=''student'') is trained on examples which are generated by a
different network (=''teacher''), then the student achieves overlap to
the teacher network. This means that the student has not only learned
 the training data but it also can classify unknown input data to
some extent - it generalizes. Using statistical mechanics, the
generalization error has been calculated exactly as a function of the
number of examples for many different scenarios and network
architectures \cite{Opper:Generalization}.

An important application of neural networks is the prediction of time series. 
There are many situations where a sequence 
of numbers is measured and one would like to know the following numbers 
without knowing the rule which produces these 
numbers \cite{Weigand}. There are powerful linear prediction 
algorithms including assumptions on external noise on the data, 
but neural networks have proven to be competitive algorithms 
compared to other known methods.

Since 1995 the statistical physics of time series prediction has been studied
\cite{Eisenstein:BG}. Similar to the static case, the 
series is generated by a well known rule - usually a different 
''teacher''--network - and the student network is trained 
on these data while moving it over the series.
We are interested in the following questions:

\begin{enumerate}
\item How well can the student network predict the 
 numbers of the series after it has been trained on part of it?
\item Has the student network achieved some knowledge 
about the rule (=network) which produced the time series?
\end{enumerate}

It seems to be straightforward to extend the analytic 
methods and results of the static classification problem to the 
case of time series prediction. The only difference 
seems to be the correlation between input vector and output bit. 
However, although many experts in this field looked 
into this problem, neither the capacity problem nor the prediction 
problem could be solved analytically up to now, even
 for the simple perceptron. Furthermore, it turned out that already 
the problem of the generation of a time series by a 
neural network is not trivial. A network can produce quasiperiodic or 
chaotic sequences, depending on the weights and transfer
functions. For some models an analytic solution has been 
derived, even for multilayer networks \cite{Kanter:Analytical}.

In this talk I intend to give an overview over 
the statistical physics of neural networks which 
generate and predict time 
series. Firstly, I discuss the capacity problem: 
Given a random sequence, what is the maximal length a perceptron can 
learn perfectly? Secondly, in Section 3 a network generating 
binary or continuous sequences is introduced and analysed. Thirdly, the 
prediction of quasiperiodic and chaotic sequences 
is investigated in Section 4. In Section 5  it is shown that  
for any prediction algorithm a sequence 
can be constructed for which this algorithm completely fails. 
Section 6 considers the problem of a set of neural 
networks which learn from each other. This scenario is applied 
to a simple economic model, the minority game. Finally, in Section 8 it 
is shown that a simple perceptron can be trained to predict a 
sequence of bits entered by the reader, even if he/she 
tries to generate random bits.

\section{Learning from random sequences}
\begin{figure}[ht]
\begin{center}
\includegraphics[width=.6\textwidth]{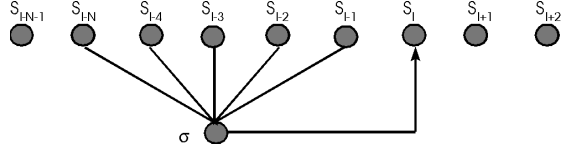}
\end{center}
\caption[]{A perceptron moves over a time series}
\label{fig1}
\end{figure}

A neural network learns from examples. 
In the case of time series prediction the examples are defined
by moving the network over the sequence, as shown in Fig.\ref{fig1}.

Let us consider the simplest possible neural network, the perceptron. 
It consists of an $N$--dimensional weight vector 
$\underline{w} = (w_1, ..., w_N)$ and a transfer function 
$\sigma = f\left(\frac{1}{N} \underline{w} \cdot 
\underline{S}\right)$.  $\underline S$ is the input vector, 
given by the sequence. We mainly consider two transfer 
functions, the Boolean and the continuous perceptron:

\begin{eqnarray}
\label{eins}
\sigma & = & \mbox{sign} (\underline{w} \cdot \underline{S});\\
\label{zwei} 
\sigma & = & \tanh \left( \frac{\beta}{N} \ \underline{w} 
     \cdot \underline{S} \right).
\end{eqnarray}

$\beta$ is a parameter giving the slope of the linear part of 
the transfer function in the continuous case, $f(x) \simeq 
\beta x + O(x^3)$.

The aim of our network is to learn a given sequence 
$S_0, S_1, S_2,...\; $.  This means that the network should find - by 
some simple algorithms - a weight vector $\underline{w}$ 
with the property
\begin{equation}
\label{drei}
S_t = f\left( \frac{1}{N} \sum\limits^{N}_{j = 1} \; w_j S_{t-j} \right)
\end{equation}
for all time steps $t$. For the Boolean function 
Eq.(\ref{eins}) 
this set of equations becomes a set of inequalities
\begin{equation}
\label{vier}
S_t \sum\limits^{N}_{j=1} \; w_j S_{t-j} > 0
\end{equation}
for all $t$. If the bits $S_t$ in Eq.(\ref{vier})
are random, $S_t \in \{+1, -1\}$, instead of taken from the time 
series then the inequalities (\ref{vier}) have a solution if the
number of inequalities is smaller than $2N$ (with 
probability one in the limit $N \rightarrow\infty$).
This is the famous result which was found by Schl\"{a}fli in about 1850 
and was calculated using replica theory by Gardner 140 years later
\cite{Schlaeffli:1850,Gardner:89}.

What happens if the bits $S_t$ are not independent 
but taken from a random time series? 
Let us assume that we arrange $P= \alpha N$ bits $S_t$ on a ring or, 
equivalently, look at $P$ random bits periodically repeated. 
For this case we ask the question: How long is the typical 
sequence which a Boolean perceptron can learn perfectly?

Up to now there is no analytical solution of Eq.(\ref{vier}) 
for this scenario, although several experts in this field have 
tried to solve this problem. However, detailed numerical
simulations show that it is harder to learn a random sequence 
than random patterns: the maximal length of the sequence 
is $P/N = \alpha_c \simeq 1.7$ \cite{Schroeder:Capacity}, which 
should be compared with 
$\alpha_c = 2$ for  random patterns. Obviously tiny 
correlations between input vectors and output bits make the 
problem harder to learn for a perceptron.

\section{Generating sequences}
%\label{SecGenSeq}
%
In the previous section the perceptron learned a short random sequence
exactly.  Consequently it also can predict it, without errors.  If a
neural network is able to {\it predict }a given time series 
it can also {\it generate} the same series.
Generating means, according to Fig.\ref{fig1}, the network takes the
last $N$ numbers of the sequence, calculates a new number and moves
one step to the right.  Repeating this procedure generates a sequence
$S_0, S_1, S_2 \dots$ given by Eq.(\ref{drei}).

Therefore it is interesting to study the structure 
of sequences generated by a neural network. Here we discuss the case 
of {\it fixed} weights $\underline{w}$, only. Adaptive
weights are considered in sections \ref{SecAntipred} to \ref{SecHuman}. 

Numerical simulations show that for random weights $\underline{w}$ and
random initial states $\underline{S}$ the sequence has a transient
initial part and finally runs into a one of several possible cycles.
The structure of these cycles is related to the maxima of the Fourier
spectrum of the weights $w_1, ..., w_N$.  Hence it is important to
understand the sequence generated by a single Fourier component
\begin{equation}
\label{fuenf}
w_j = \cos \left( 2 \pi K \frac{j}{N} + \pi \phi \right).
\end{equation}
$K$ is an integer frequency and $\phi \in [-1, 1]$ 
a phase of the weight vector. For a continuous perceptron we are 
looking for a solution $S_0, S_1 \dots$ of an infinite number of equations
\begin{equation}
\label{sechs}
S_t = \tanh \left[ \frac{\beta}{N} \; \sum^{N}_{j=1} 
  \cos \left( 2 \pi K \frac{j}{N} - \pi \phi \right) S_{t-j} 
\right].
\end{equation}
For this case an analytic solution could be derived \cite{Kanter:Analytical}. 
For small values of $\beta$ the attractor is zero, the 
sequence relaxes to $S_t = 0$. However, above a critical 
value of $\beta$ which is independent of the frequency $K$, a 
nonzero attractor exists; close to $\beta_c$ it is given by 
\begin{equation}
\label{sieben}
S_t = \tanh \left[ A (\beta) \cos \left( 2\pi (K + \phi) 
\frac{t}{N} \right) \right].
\end{equation}
The amplitude $A(\beta)$ increases continuously from zero above a
critical value
\begin{equation}
\label{acht}
\beta > \beta_c = 2 \frac{\pi \phi}{\sin (\pi \phi)}.
\end{equation}
Therefore, the attractor of the sequence is a {\it quasiperiodic}
 cycle with a frequency $K + \phi$. The phase $\phi$ of the 
weights shifts the frequency of the sequence - a result which 
is not easy to understand without calculating it.

\begin{figure}[ht]
\begin{center}
\includegraphics[width=.6\textwidth]{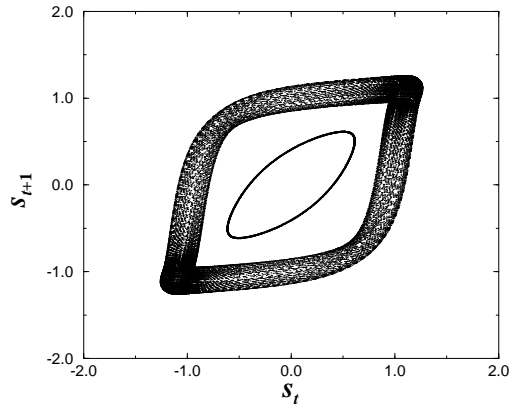}
\end{center}
\caption[]{Attractor of a network with two hidden units}
\label{fig2}
\end{figure}

For a multilayer network the situation is similar: Each hidden unit
can contribute a quasiperiodic component to the sequence, which has
its own critical point.  Increasing $\beta$, more and more components
are activated.  This is shown in Fig.\ref{fig2} for a network with
two hidden units: For small values of the parameter $\beta$ the
quasiperiodic attractor is one--dimensional, for large $\beta$ both
compoments are activated yielding a two--dimensional attractor as shown by the
return map $S_{t+1} (S_t)$. The attractor dimension is limited by the
number of hidden units \cite{Kanter:AttrDim}.

 If the transfer function is discrete, Eq.(\ref{eins}), the 
situation is more complex \cite{Eisenstein:BG,Schroeder:Cycles}
In this case we 
obtain a bit generator whose cycle length is limited by $2^N$. 
However, numerical simulations show that the spectrum of 
cycle lengths has a much lower bound, namely the value $2N$, at least for single component
weights with $|\phi| < 1/2$. 
After a transient part the bit sequence $S_t$ follows the 
equation
\begin{equation}
\label{neun}
S_t = \mbox{sign} \left[ \cos \left( 2 \pi (K + \phi)\right)  
\frac{t}{N} \right]. 
\end{equation}
But the sequence cannot follow this equation forever; namely if a
window \linebreak[4] $(S_{t-1}, ... , S_{t-N})$  appeares a second time, the perceptron
has to repeat the sequence. Numerical calculations show that Eq.(\ref{neun}), in addition to this
condition,  produces cycles shorter than $2N$. It remains a
challenge to show this result analytically.

\begin{figure}[ht]
\begin{center}
\includegraphics[width=.6\textwidth]{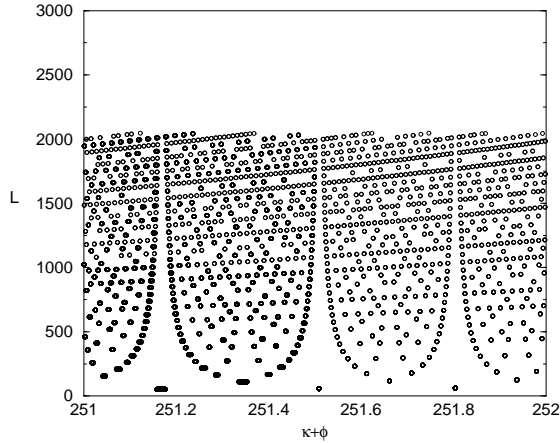}
\end{center}
\caption[]{Cycle lengths of the bit generator with Cosine weights (\ref{fuenf}). 
The perceptron has $N=1024$ input components.}
\label{fig3}
\end{figure}

Fig.\ref{fig3} shows the cycle length $L(\phi) $ of the bit generator
with weights (\ref{fuenf}). This rather complex figure has a simple origin, it just shows the
properties of rational numbers. An integer multiple of the wavelength
$\lambda$ given by (\ref{neun})
\begin{equation}
\label{zehn}
\lambda = \frac{N}{K+ \phi}
\end{equation}
has to fit into the cycle
\begin{equation}
\label{elf}
L = n \cdot \lambda.
\end{equation}
Hence $\lambda$ has to be a rational. The pattern $L(\phi)$ shown in Fig.3 
turns out to be the numerator as a function of its rational basis. 
However, this does not explain why this picture is cut 
for $L > 2N$.

Up to now we have discussed quasiperiodic sequences, only. But 
time series occurring in applications are in general more complex. 
Therefore we are interested in the question: Can 
a neural network generate a time series with a more complex 
power spectrum than a single peak and its higher harmonics?

It turns out that a multilayer network cannot generate a 
sequence with an arbitrary power spectrum. To generate a sequence 
with autocorrelations which decay as a power law, one needs 
a fully connected asymmetric network, a more complex 
architecture than a feedforward network \cite{Priel:Correlated}.

However, a simple perceptron can generate a chaotic sequence. 
When the weights have a bias,
\begin{equation}
\label{zwoelf}
b = \frac{1}{N} \sum\limits_i w_i >0,
\end{equation}
then there are tiny regions in the $(\beta, b)$--plane 
where a chaotic sequence has been observed numerically 
\cite{Priel:Long-term}. Such a scenario has 
been called {\it fragile chaos}. The fractional dimension of such a 
chaotic sequence is between one and two, and in the vicinity of 
chaotic parameters $(\beta, b)$ there is always a 
parameter set with a quasiperiodic sequence. 

This situation is different for a nonmonotonic transfer function. If the
function $\tanh (x)$ in (\ref{zwei}) is replaced by $\sin(x)$ there are
large compact regions in the parameter space where the sequence is
chaotic with a large fractal dimension of the order of $N$. Neural
networks with nonmonotonic transfer functions yield high dimensional
{\it stable chaos} \cite{Priel:Long-term,Priel:Robust}. In this case
the attractor dimension can be tuned by the parameter $\beta$ between
the values one and $N$.

\section{Prediciting time series}
\label{SecPred}
If a neural network cannot generate a given sequence of numbers, it
cannot predict it with zero error. But this is not the whole story.
Even if the sequence has been generated by an (unknown) neural network
(the teacher), a different network (the student) can try to learn and
to predict this sequence. In this context we are interested in two
questions:

\begin{enumerate}
\item When a student network with the identical architecture as the
  teacher one is trained on the sequence, how does the overlap between
  student and teacher develop with the number of training examples (=
  windows of the sequence)?
\item After the student network has been trained on a 
part of the sequence how well can it predict the sequence several 
steps ahead?
\end{enumerate}
Recently these questions have been investigated numerically 
for the simple perceptron \cite{Freking}. We have to distinguish 
several scenarios:
\begin{enumerate}
\item Boolean versus continuous perceptron
\item On--line versus batch learning
\item Quasiperiodic versus chaotic sequence.
\end{enumerate}
In all cases we consider only the stationary part of a 
sequence which was generated by a perceptron. The student network 
is trained on the stationary part only, not on the transient.

First we discuss the Boolean perceptron of size $N$ which has
generated a bit cycle with a typical length $L<2N$. The teacher
perceptron has random weights with zero bias, and the cycle is related
to one component of the power spectrum of the weights. The student
network is trained using the perceptron learning rule:
\begin{eqnarray}
\label{dreizehn}
\Delta w_i & =  \frac{1}{N} S_t \; S_{t-i} & \; \;\mbox{if} \; \; S_t 
\sum\limits^{N}_{j=1} w_j S_{t-j} <0; \nonumber \\ 
\Delta w_i & = 0 \mbox{~~else}.
\end{eqnarray}

For this algorithm there exists a mathematical theorem
\cite{Hertz:NeuralComp}: If the set of examples can be generated by
some perceptron then this algorithm stops, i.e.  it finds one out of
possibly many solutions.  Since we consider examples from a bit
sequence generated by a perceptron, this algorithm is guaranteed to
learn the sequence perfectly. On--line and batch training are
identical, in this case.

The network is trained on the cycle until the training error is zero.
Hence the student network can predict the stationary sequence
perfectly. Surprisingly, it turns out that the overlap between student
and teacher is small, in fact it is zero for infinitely large
networks, $N\rightarrow \infty$. The network learns the projection of
the teacher's weight vector onto the sequence, but not the complete
vector. It behaves like a filter selecting one of the components of
the power spectrum of the weights. Although it predicts the sequence
perfectly, it does not gain much information on the rule which
generates this sequence.

This situation seems to be different in the case of a continuous 
perceptron. Inverting Eq.(\ref{drei}) for a monotonic 
transfer function $f(x)$ gives $N$ linear equations for $N$ 
unknowns $w_i$. If the stationary part of the sequence is 
either quasiperiodic or chaotic, 
all patterns are different and the batch training, using $N$ windows,
leads to perfect learning.

This holds true for a chaotic time series.  However, for a
quasiperiodic one (Eq.(\ref{sieben})) the patterns are almost linearly
dependent, yielding an ill--conditioned set of linear equations. Without
the $ \tanh (x)$ in Eq.(\ref{sieben}), one would obtain a
two--dimensional space of patterns; with the nonlinearity one obtains small
contributions in the other $N-2$ dimensions of the weight space.
Nevertheless, depending on the parameter $\beta $, even professional
computer routines sometimes do not succeed in solving Eq.(\ref{drei})
for quasiperiodic patterns generated by a teacher perceptron.

How does this scenario show up in an on--line training algorithm 
for a continuous perceptron? If a
quasiperiodic sequence is learned step by step without iterating
previous steps, using gradient descent to update the weights,
\begin{equation}
\label{vierzehn}
\Delta w_i = \frac{\eta}{N} (S_t - f(h)) \cdot f^{\prime} (h) \cdot
S_{t-i} \; \; \mbox{with} \; \; h= \beta \sum\limits^{N}_{j=1} w_j S_{t-j}
\end{equation} 
then one can distinguish two time scales (time = number of training steps):
\begin{enumerate}
\item A fast one increasing the overlap between teacher and student to
  a value which is still far away from the value one which corresponds
  to perfect agreement.
\item A slow one increasing the overlap very slowly. 
Numerical simulations for millions times $N$ training steps yielded 
an overlap which was still far away from the value one.
\end{enumerate}

Although there is a mathematical theorem on stochastic optimization
which seems to guarantee convergence to perfect success
\cite{StochOpt}, our on--line algorithm cannot gain much information
about the teacher network. It would be interesting to know how these
two time scales depend on the size of the system. In addition we cannot exclude that
there exist on-line algorithms which can learn our ill-conditioned problem
in short times.

This is completely different for a chaotic time series generated by a
corresponding teacher network with $f(x)=sin(x)$. It turns out that the
chaotic series appears like a random one: After a number of training steps of
the order of $N$ the overlap relaxes exponentially fast to perfect
agreement between teacher and student.

Hence, after training the perceptron with a number of examples 
of the order of $N$ we obtain the two cases:
For a quasiperiodic sequence the student has not obtained 
much information about the teacher, while for a chaotic 
sequence the student's weight vector comes close to the one of the
teacher. One important question remains: How well can the 
student predict the time series?

\begin{figure}[ht]
\begin{center}
\includegraphics[width=.6\textwidth]{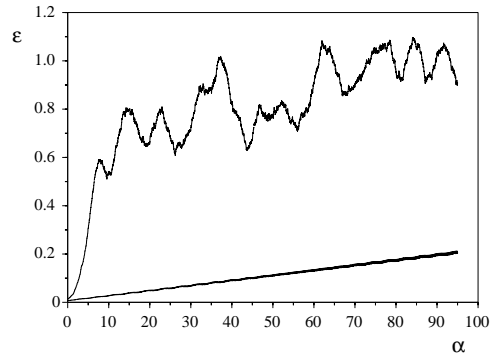}
\end{center}
\caption[]{Prediction error as a function of 
time steps ahead, for a quasiperiodic (lower) and chaotic (upper) series.}
\label{figpred}
\end{figure}

Fig.\ref{figpred} shows the prediction error as a function of the
time interval over which the student makes the predictions. The
student network which has been trained on the quasiperiodic sequence
can predict it very well. The error increases linearly with the size
of the interval, even predicting $10N$ steps ahead yields an error of
about 10\% of the total possible range. On the other side, the student
trained on the chaotic sequence cannot make predictions. The prediction
error increases exponentially with time; already
after a few  steps the error corresponds to random guessing, $ \epsilon
\simeq 1$.

In summary we obtain the surprising result:
\begin{enumerate}
\item A network trained on a quasiperodic sequence does not 
obtain much information  about the teacher network which generated 
the sequence. But the network can predict this sequence 
over many (of the order of $N$) steps ahead.
\item A network trained on a chaotic sequence obtains almost 
complete knowledge about the teacher network. But this 
network cannot make reasonable predictions on the sequence.
\end{enumerate}

It would be interesting to find out whether this result 
also holds for other prediction algorithms, such as multilayer networks.

\section{Predicting with 100\% error}
\label{SecAntipred}
Consider some arbitrary  prediction  algorithm. It may contain all the 
knowledge of mankind, many experts may have developed it. Now 
there is a bit sequence $S_1, S_2 ...$ and the algorithm
 has been trained on the first $t$ bits $S_1, ..., S_t$. Can it 
predict the next bit $S_{t+1}$? Is the prediction error,
 averaged over a large $t$ interval, less than 50\%?

If the bit sequence is random then every algorithm will 
give a prediction error of 50\%. But if there are some correlations 
in the sequence then a clever algorithm should be able 
to reduce this error. In fact, for the most powerful algorithm one 
is tempted to say that for {\it any} sequence it should
 perform better than 50\% error. However, this is not 
true \cite{Zhu:Antipred}.
To see this just generate a sequence $S_1, S_2, S_3, ...$ 
using the following algorithm: \\ 

\centerline{\fbox{
\parbox{10cm}{
Define $S_{t+i} $ to be the opposite of the prediction 
of the algorithm which has been trained on 
$S_1, \dots, S_t$ .}}}

\vspace{4mm}
Now, if the same  algorithm is trained on this sequence, it will 
always predict the following bit with 100\% error. Hence there 
is no general prediction machine; to be successful for a
class of problems the algorithm needs some preknowledge about 
it. 

The Boolean perceptron is a very simple prediction algorithm
for a bit sequence, in particular with the on--line 
training algorithm (\ref{dreizehn}). 
How does the bit sequence look like for which the perceptron 
completely fails?

Following (\ref{dreizehn}) we just have to take the negative value
\begin{equation}
\label{sechzehn}
S_t = - \mbox{sign} \left( \sum\limits^N_{j=1} \; w_j S_{t-j} \right)
\end{equation}
and then train the network on this new bit
\begin{equation}
\label{siebzehn}
\Delta w_j = + \frac{1}{N} S_t \; S_{t-j}.
\end{equation}
The perceptron is trained on the opposite (= negative) of its own
prediction. Starting from (say) random initial states $S_1, \dots,
S_N$ and weights $\underline{w}$, this procedure generates a sequence
of bits $S_1, S_2, \dots S_t, \dots$ and of vectors $\underline{w},
\underline{w}(1), \underline{w}(2), \linebreak \dots \underline{w}(t), \dots$ as
well. Given this sequence and the same initial state, the perceptron
which is trained on it yields a prediction error of 100\%.

It turns out that this simple algorithm produces a rather 
complex bit sequence which comes close to a random one. After a 
transient time the weight vector $\underline{w}(t)$
 performs a kind of random walk on a $N$--dimensional 
hypersphere. The bit sequence runs to a cycle whose average length
 $L$ scales exponentially with $N$,
\begin{equation}
\label{achtzehn}
L \simeq 2.2^N.
\end{equation}
The autocorrelation function of the sequence shows complex properties: 
It is close to zero up to $N$, oscillates between $N$ and 
$3N$ and it is similar to  random noise for larger distances. Its entropy 
is smaller than the one of a random sequence since the 
frequency of some patterns is suppressed. Of course, 
it is not random since the prediction error is 100\% instead of 50\% 
for a random bit sequence.

When a second perceptron (=student) with different initial state
$\underline{w}$ is trained on such an antipredictable sequence
generated by Eq.(\ref{dreizehn})  it can perform somewhat better
than the teacher: The prediction error goes down to about 78\% but it
is still larger than 50\% for random guessing. However, the student
obtains knowledge about the teacher: The angle between the two weight
vectors relaxes to about 45 degrees.

\section{Learning from each other}
\label{SecLearn}
In the previous section we have discussed a neural network 
which learns from itself. But more interesting may be the 
scenario where several networks are interacting, learning 
from each other. After all, our living world consists of 
interacting adaptive systems and recent methods of computer 
science use interacting agents to solve complex problems. 
Here we consider a simple system of interacting perceptrons 
as a first example to develop a theory of cooperative 
behaviour of adaptive agents.

Consider $K$ Boolean perceptrons, each of which has an 
$N$--dimensional weight vector $\underline{w}^{\nu}, \nu = 1, \dots, 
K$. Each perceptron is receiving the same input vector
 $S_1, \dots, S_N$ and produces its own output bit
\begin{equation}
\label{neunzehn}
\sigma^{\nu} = \mbox{sign} (\underline{w}^{\nu} \cdot \underline{S})
\end{equation}
Now these networks receive information from their neighbours 
in a ring--like topology: Perceptron $\underline{w}^{\nu}$ is 
trained on the output $\sigma^{\nu - 1}$ of perceptron 
$\underline{w}^{\nu - 1}$, and $\underline{w}^1$ is trained on 
$\sigma^K$. Training is performed keeping the length of the 
weight vectors fixed:
\begin{equation}
\label{zwanzig}
{\underline{w}^{\nu}}(t+1) = 
\frac{{\underline{w}^{\nu}}(t) + (\eta/N) \sigma^{\nu-1} \underline{S}}
{|{ \underline{w}^{\nu}}(t) + (\eta/N) \sigma^{\nu-1} \underline{S}|}
\end{equation}
The learning rate $\eta$ is a parameter controlling the speed 
of learning.

This problem has been solved analytically in the 
limit $N \rightarrow \infty$ \cite{Metzler:Interact} 
for random inputs. The system 
relaxes to a stationary state, where the angles 
$\theta_{\nu \mu}$ (or overlaps) between different agents take a fixed 
value. For small learning rate $\eta$ all of 
these angles are small, i.e. there is good agreement between the 
agents. But more surprising: The state of the 
system is completely symmetric, there is only one common angle $\theta = 
\theta_{\nu \mu}$ 
between all pairs of networks. The agents do not
 recognize the clockwise flow of information.

Increasing the lerning rate $\eta$ the common angle 
$\theta$ increases, too. With larger learning steps each agent tends to 
have an opinion opposite to  all of its colleagues. 
But, due to the symmetry, there is maximal possible angle given by
\begin{equation}
\label{einundzwanzig}
\cos{\theta} = -\frac{1}{K-1}.
\end{equation}
In fact, increasing $\eta$ the system arrives at this maximal angle at some critical value $\eta_c$. For larger value of 
$\eta > \eta_c$ the system undergoes a phase transition: The complete symmetry is broken, but the symmetry of the ring is 
still conserved:
\[
\theta_1 = \theta_{\nu +1, \nu} \; \; , \; \; \theta_2 = 
\theta_{\nu +2, \nu}, ...
\]
For $K$ agents there are  $(K-1)/2$ values of $\theta_i$ possible
if $K$ is odd, and $K/2-1$ values for even $K$.

This is a simple - but analytically solvable - example of a 
system of interacting neural networks. We observe a symmetry 
breaking transition when increasing the learning rate. However, 
this system does not solve any problem. In the following 
section we will extend this scenario to a case where indeed 
neural networks interact to solve a special problem, the 
minority game.

\section{Competing in the minority game}
\label{SecMin}

Recently a mathematical model of economy receives a lot 
of attention in the community of statistical physics 
\cite{Econophys}. 
It is a simple model of a closed market: There are 
$K$ agents who have to make a binary decision 
$\sigma^{\nu} \in \{+1, -1\}$ at each time step. All of 
the agents who belong to the minority gain one point, the majority 
has to pay one point (to a cashier which always wins). 
The global loss is given by
\begin{equation}
\label{zweiundzwanzig}
G = \left| \sum\limits^{K}_{\nu =1} \; \; \sigma^{\nu} \right|
\end{equation}
If the agents come to an agreement before they make a new 
decision, it is easy to minimize $G: (K-1)/2$ agents have to 
choose $+1$, then $G =1$. However, this is not the rule 
of the game, the agents are not allowed to cooperate. Each agent 
knows only the history of the minority decision, 
$S_1, S_2, S_3, \dots$, but otherwise he/she has no information. Can the 
agent find an algorithm to maximize his/her profit?

If each agent makes a random decision, then $G$ is of the order of $\sqrt{K}$. It is not easy to find algorithms which 
perform better than random \cite{Challet:Theory,Reents:Stochastic}. 

Here we use a perceptron for each agent to make a decision based 
on the past $N$ steps $\underline{S} = (S_{t-N}, ..., 
S_{t-1})$ of the minority decision. The decision of agent 
$\underline{w}^{\nu}$ is given by 
\begin{equation}
\label{dreiundzwanzig}
\sigma^{\nu} = \mbox{sign} (\underline{w}^{\nu} \; \underline{S}).
\end{equation}
After the bit $S_t$ of the minority has been determined, each 
perceptron is trained on this new example $(\underline{S}, S_t)$,
\begin{equation}
\label{vierundzwanzig}
\Delta \underline{w}^{\nu} = \frac{\eta}{N} \; \; S_t \; \underline{S}.
\end{equation}
This problem could be solved analytically \cite{Metzler:Interact}. 
The average global loss for $\eta \rightarrow 0$ is given by
\begin{equation}
\label{fuenfundzwanzig}
\langle G^2 \rangle = (1-2/\pi) K \simeq 0.363 \; K.
\end{equation}
Hence, for small enough learning rates the system of interacting 
neural networks performs better than random decisions. A 
pool of adaptive perceptrons can organize itself to yield a 
successful cooperation.

\vspace{1cm}

\section{Predicting human beings}
\label{SecHuman}

As a final example of a perceptron predicting a bit sequence we 
discuss a real application. Assume that the bit sequence 
$S_0, S_1, S_2, ...$ is produced by a human being. Now a simple 
perceptron (\ref{eins}) with on--line learning 
(\ref{dreizehn}) takes the last $N$ bits and makes a prediction for 
the next bit. Then the network is trained on the new true bit, 
which afterwards appears as part of the input for the following
 prediction.
 
 Eq.(\ref{dreizehn}) is a simple deterministic equation describing the
 change of weights according to the new bit and the past $N$ bits. Can
 such a simple equation foresee the reaction of a human being? On the
 other side, if a person can calculate or estimate the outcome of Eq.
 (\ref{dreizehn}), then he/she can just do the opposite, and the
 network completely fails to predict.

To answer these questions we have written a little C program
 which receives the two bits 0 and 1 from the keyboard 
\cite{KR:98}. The program needs two fields {\tt neuron} 
and {\tt weight} which contain the variable $S_i$ 
and $w_i$, respectively. Here are the main steps:
\begin{enumerate}
\item Repeat:\\
{\tt while (1)  \{ }
\item Calculate the vector product $\underline{w} \, \underline{S}$:\\
{\tt for (h=0; i=0; i<N; i++)  h+=weight[i]*neuron[i];}
\item Read a new bit:\\
{\tt if(getchar()=='1')  \ \ input=1; else input =-1;}
\item Calculate the prediction error:\\
{\tt if(h*input<0) \{error ++;}
\item Train:\\
{\tt for(i=0; i<N; i++)  weight[i]+=input* neuron[i]/(double)N;\} }
\item Shift the input window:\\
{\tt for(i=N-1; i>0; i--)  neuron[i]=neuron[i-1]; neuron[0] =input; \} }
\end{enumerate}
A graphical version of this program can be accessed over the internet:\\
\centerline{ http://theorie.physik.uni-wuerzburg.de/\~{ }kinzel}

Now we ask a person to generate a bit sequence for which 
the prediction error of the network is high. We already know 
from section 2 what happens if the candidate produces a rhythm: 
if its length is smaller than $1.7 N$ the perceptron can 
learn it perfectly, without errors. Hence the candidate should either produce 
random numbers which give 50\% errors or he/she should try 
to calculate the prediction of the perceptron,  in this case 
an error higher than 50\% is possible.

We have tested this program on students of our class.  Each student
had to send a file with one thousand bits, generated by hand. It turns
out that on average the network predicts with an error of about 35\%.
The distribution of errors is broad with a range between 20\% and
50\%.  Apparently, a human being is not a good random number generator.  The
simple perceptron (\ref{eins}) and (\ref{dreizehn}) succeeds in
predicting human behaviour!

Some students submitted sequences with 50\% error. 
It was obvious - and later confessed - that they used random number 
generators, digits of $\pi$, the logistic map, etc. instead of 
their own fingers. One student submitted a sequence with 100\% 
error. He was the supervisor of our computer system, knew the 
program and submitted the sequence described in section 5.

\section{Summary}
\label{SecSummary}

The theory of time series generation and prediction is a 
new field of statistical physics. The properties of perceptrons, 
simple single--layer neural networks being trained on sequences 
which were produced by other perceptrons, have been studied. 
A random bit sequence is more difficult to learn perfectly than 
random uncorrelated patterns. An analytic solution of 
this capacity problem is still missing.

A multilayer network can be used to generate time series. 
For the continuous transfer function an analytic solution of 
the stationary part of the sequence has been found. The 
sequence has a dimension which is bounded by the number of hidden 
units. It is not completely clear yet how to extend this
solution to the case of a Boolean perceptron generating a bit 
sequence. For nonmonotonic transfer functions the network 
generates a chaotic sequence with a large fractal dimension.

A perceptron which is trained on a quasiperiodic sequence 
can predict it very well, but it does not obtain much 
information on the rule generating the sequence. On the 
other side, for a chaotic sequence the overlap between student 
and teacher is almost perfect, but prediction of the 
sequence is not possible.

For any prediction algorithm there is a sequence for 
which it completely fails. For a simple perceptron such a sequence 
is rather complex, with huge cycles and low autocorrelations.
Another perceptron which is trained on such a sequence 
reduces the prediction error from 100\% to 78\%  
and obtains overlap to the generating network.

When perceptrons learn from each other, the system relaxes 
to a symmetric state. Above a critical learning rate there is a 
phase transition to a state with lower symmetry.

A system of interacting neural network can develop 
algorithms  for the minority game, a model of a closed economy of 
competing agents.

Finally it has been demonstrated that human beings 
are not good random number generators. Even a simple perceptron can 
predict the bits typed by hand with an error of less than 50\%.\\

{\bf Acknowledgement:}

The author appreciates comments of Ido Kanter, Richard Metzler, 
Ansgar Freking and Michael Biehl.

%INDEX%%%%%%%%%%%%%%%%%%%%%%%%%%%%%%%%%%%%%%%%%%%%%%%%%%%%%%%%%%%%%%%
% Please check with the editor of your book whether he plans to
% include a "mutual" subject index - if so, please code your entries
% in the standard syntax. For your own purposes you may print your
% "personal" index by using the following commands:
%
%\clearpage
%\addcontentsline{toc}{section}{Index}
%\flushbottom
%\printindex
%%%%%%%%%%%%%%%%%%%%%%%%%%%%%%%%%%%%%%%%%%%%%%%%%%%%%%%%%%%%%%%%%%%%%


\begin{thebibliography}{8.}
\addcontentsline{toc}{section}{References}

\bibitem{Hertz:NeuralComp} J. Hertz, A. Krogh, and R. G. Palmer:
\emph{Introduction to the Theory of Neural Computation}, 
(Addison Wesley, Redwood City, 1991)

\bibitem{Engel:00} A. Engel, and  C. Van den Broeck:
\emph{Statistical Mechanics of Learning}, (Cambridge University Press, 2000)

\bibitem{Kinzel:Rev} W. Kinzel:
 \emph{Statistical physics of neural networks}, Comp. Phys. Comm.
\textbf{121}, 86-93 (1999)

\bibitem{BiehlC} M. Biehl and N. Caticha: 
Statistical Mechanics of On-line Learning and Generalization,
\emph{The Handbook of Brain Theory and Neural Networks}, ed. by M. A. Arbib (MIT Press, Berlin 2001)

\bibitem{Opper:Generalization} M. Opper and W. Kinzel: 
Statistical Mechanics of Generalization,
\emph{Models of Neural Networks III}, ed. by  E. Domany and J.L. van Hemmen and K. Schulten,
151-209  (Springer Verlag, Heidelberg 1995)

\bibitem{Weigand} A. Weigand and N. S. Gershenfeld:
\emph{Time Series Prediction}, Santa Fe, (Addison Wesley, 1994)

\bibitem{Eisenstein:BG} E. Eisenstein and I. Kanter and D.A. Kessler and W. Kinzel:
\emph{Generation and Prediction of Time Series by a Neural
Network}, Phys. Rev. Letters  
\textbf{74} 1, 6-9 (1995)

\bibitem{Kanter:Analytical} I. Kanter and D.A. Kessler and A. Priel and E. Eisenstein:
\emph{Analytical Study of Time Series Generation by Feed-Forward
Networks}, 
Phys. Rev. Lett.   
\textbf{75} 13, 2614-2617 (1995)

\bibitem{Schlaeffli:1850} L. Schlaefli:
\emph{Theorie der Vielfachen Kontinuitaet}, 
(Birkhaeuser, Basel 1857)   

\bibitem{Gardner:89} E. Gardner:
\emph{The space of interactions in neural network models }, 
J. Phys. A   
\textbf{21}, 257 (1988)

\bibitem{Schroeder:Capacity} M. Schr{\"{o}}der and W. Kinzel and I. Kanter:
\emph{Training a perceptron by a bit sequence: storage capacity}, 
J. Phys. A   
\textbf{29}, 7965 (1996)

\bibitem{Kanter:AttrDim} L. Ein-Dor and I. Kanter:
\emph{Time Series Generation by Multi-layer networks}, 
Phys. Rev. E   
\textbf{57}, 6564 (1998)

\bibitem{Schroeder:Cycles} M. Schr{\"{o}}der and W. Kinzel:
\emph{Limit cycles of a perceptron}, J. Phys. A 
\textbf{31}, 9131-9147 (1998)
      
\bibitem{Priel:Correlated} A. Priel and I. Kanter:
\emph{Learning and generation of long-range correlated sequences}, 
Phys. Rev. E, in press

\bibitem{Priel:Long-term} A. Priel and I. Kanter:
\emph{Long-term properties of time series generated by 
a perceptron with various transfer functions}, Phys. Rev. E, 
\textbf{59} 3, 3368-3375 (1999)

 \bibitem{Priel:Robust} A. Priel and I. Kanter:
\emph{Robust chaos generation by a perceptron}, 
 Europhys. Lett. \textbf{51}, 244-250 (2000)

 \bibitem{Freking} A. Freking and W. Kinzel and I. Kanter:
\emph{unpublished}

\bibitem{StochOpt} C. M. Bishop:
\emph{Neural Networks for Pattern Recognition}
(Oxford University Press, New York 1995) 

\bibitem{Zhu:Antipred} H. Zhu and W. Kinzel:
\emph{Anti-Predictable Sequences: Harder to Predict Than A Random Sequence}, 
Neural Computation   
\textbf{10}, 2219-2230 (1998)

\bibitem{Metzler:Interact} W. Kinzel, R. Metzler, I. Kanter, \emph{Dynamics of interacting
neural networks}, J. Phys. A \textbf{33} L141-L147 (2000);\\
 R. Metzler and W. Kinzel and I. Kanter:
\emph{Interacting Neural Networks}, 
Phys. Rev. E   
\textbf{62} 2, 2555 (2000)

\bibitem{Econophys} Econophysics homepage:
http://www.unifr.ch/econophysics/

\bibitem{Challet:Theory} D. Challet and M. Marsili and R. Zecchina:
\emph{Statistical Mechanics of Systems with Heterogeneous  Agents: Minority Games}, 
Phys. Rev. Lett.  
\textbf{84} 8, 1824-1827 (2000)

\bibitem{Reents:Stochastic} G. Reents and R. Metzler and W. Kinzel:
\emph{A New Stochastic Strategy for the Minority Game}, 
 cond-mat/0007351 (2000)

\bibitem{KR:98} W. Kinzel and G. Reents: 
\emph{Physics by computer}, (Springer-Verlag, Heidelberg 1998)

          
\end{thebibliography}
\end{document}